\documentclass[a4paper,11pt]{article}
\usepackage{pos}
\usepackage{physics}
\usepackage{braket}
\usepackage{booktabs}
\usepackage{arydshln} 
\usepackage{hhline}

\title{Study on the $P$-wave form factors contributing to $ B_s $ to $D_s$ inclusive semileptonic decays from lattice simulations}
\ShortTitle{$P$-wave form factors of $ B_s $ to $D_s$ semileptonic decays}

\author*[a]{Z. Hu}
\author[b]{A. Barone}
\author[c,d]{A. Elgaziari}
\author[a,e]{S. Hashimoto}
\author[c,d,f]{A. J\"uttner}
\author[a,e]{T. Kaneko}
\author[a]{R. Kellermann}

\affiliation[a]{Theory Center, Institute of Particle and Nuclear Studies, High Energy Accelerator Research Organization (KEK), Tsukuba 305-0801, Japan}

\affiliation[b]{PRISMA+ Cluster of Excellence \& Institut f\"ur Kernphysik, Johannes-Gutenberg-Universit\"at Mainz, D-55099 Mainz, Germany}

\affiliation[c]{School of Physics and Astronomy, University of Southampton, Southampton SO17 1BJ, United Kingdom}

\affiliation[d]{STAG Research Center, University of Southampton, Southampton SO17 1BJ, UK}

\affiliation[e]{School of High Energy Accelerator Science, The Graduate University for Advanced Studies (SOKENDAI, Ibaraki 305-0801, Japan}

\affiliation[f]{CERN, Theoretical Physics Department, Geneva, Switzerland}

\emailAdd{huzhi@post.kek.jp, abarone@uni-mainz.de, A.Elgaziari@soton.ac.uk, shoji.hashimoto@kek.jp, andreas.juttner@cern.ch, takashi.kaneko@kek.jp, kelry@post.kek.jp}

\abstract{We present a pilot study on extracting the form factors of the semileptonic decay of a $ B_s $ meson to the $P$-wave $ D_s^{**} $ states from $B_s$ four-point correlators. With their inclusive nature, four-point correlators include contributions from all possible final states. From the extracted $ P $-wave form factors, we obtain numerical results for the corresponding Isgur-Wise form factors. The results suggest significant contributions from radial excitations to the Uraltsev sum rule at zero-recoil. In this pilot study, a coarse lattice of $ 24^3\times 64 $ with lattice spacing of $0.11\,\mathrm{fm}$ is used for the analysis.}

\FullConference{The 41st International Symposium on Lattice Field Theory (LATTICE2024)\\
 28 July - 3 August 2024\\
Liverpool, UK\\}

\newcommand{\ie}{\textit{i.e.}} 
 
\newcommand{\fig}[1]{Fig.~#1}

\newcommand{\tab}[1]{Tab.~#1}

\newcommand{\eq}[1]{Eq.~(#1)}
\newcommand{\eqs}[1]{Eqs.~(#1)}
\newcommand{\paper}[1]{Ref.~#1}
\newcommand{\papers}[1]{Refs.~#1}

\newcommand{\ssec}[1]{Sec.~#1}

\newcommand{\mybf}[1]{\boldsymbol{#1}}
\newcommand{\mycal}[1]{\mathcal{#1}}
\newcommand{\Bs}{{B_s}}
\newcommand{\Xcs}{{X_{cs}}}
\newcommand{\rmd}{\mathrm{d}}
\newcommand{\MBs}{M_{B_s}}

\newcommand{\VparaVpara}{V_\parallel V_\parallel}

\newcommand{\AzeroAzero}{A_0 A_0}
\newcommand{\AparaApara}{A_\parallel A_\parallel}

\newcommand{\GeV}{\mathrm{GeV}}
\newcommand{\MeV}{\mathrm{MeV}}
\newcommand{\Abs}[1]{\left|#1\right|}
\newcommand{\src}{\mathrm{src}}
\newcommand{\snk}{\mathrm{snk}}

\begin{document}
\maketitle

\section{Introduction}

Some tensions remain in flavor physics, including the well-known inconsistency between inclusive and exclusive determinations of $|V_{cb}|$ \cite{HFLAV:2022esi}. A less famous example is the puzzle concerning the $B$-meson semileptonic decay rates to excited $D$ mesons. 
According to heavy quark effective theory (HQET), the branching ratio of $ B_{(s)} $ decaying into the $P_{3/2}$-channel $ D_{(s)} $ mesons should be much larger than that to the $P_{1/2}$-channel $ D_{(s)} $ mesons, but it contradicts the experimental data of the $B$ decays. It is called the 1/2-versus-3/2 puzzle \cite{Bigi:2007qp}. This may be the source of another problem in the semileptonic $B$ decays, {\it i.e.} the gap between the sum of currently observed exclusive branching ratios of the $ b \rightarrow c $ semileptonic decays and the corresponding inclusive decay width.

The connection between inclusive and exclusive observables may be established through the four-point correlators on the lattice \cite{Hashimoto:2017wqo,Gambino:2020crt}. This work presents a pilot study to extract exclusive information from the four-point correlators of the $ B_s $ meson calculated on the lattice. In \ssec{\ref{ss:CJmuJnu}} we present the four-point correlator $ C_{J_\mu J_\nu} $ and its relation to inclusive and exclusive processes. In \ssec{\ref{ss:decomposition_CJmuJnu}}, we decompose $ C_{J_\mu J_\nu} $ using the exclusive form factors. In \ssec{\ref{ss:lattice}} we introduce the lattice setup and \ssec{\ref{ss:numerics}} contains our numerical results. We conclude this paper with discussions in \ssec{\ref{ss:discussions}}.

\section{Semileptonic decays and $ C_{J_\mu J_\nu} $} \label{ss:CJmuJnu}

We begin by elucidating the kinematics of the $ B_s $ semileptonic decays
$
    B_s \longrightarrow \Xcs l \nu_l   \label{eq:Bs_Xcs_decay}
$.
We work in the center-of-mass frame of the $ \Bs $ meson, \ie{}, $ p_{B_s} = (M_{B_s}, \mybf{0}) $. The momentum carried away by the lepton pair is $ q = p_\Bs - p_{\Xcs} = (M_\Bs - E_\Xcs , \mybf{q}) $, where $ E_\Xcs $ is the energy left for the final-state hadron.

For the inclusive semileptonic decays, the differential cross section after angular and lepton energy integrals reads
\begin{gather}
    \frac{\rmd\Gamma^{\mathrm{inc}}}{\rmd \mybf{q}^2} \propto G_F^2 \abs{V_{cb}}^2 \int \rmd E_\Xcs \, W^{\mu\nu}(\mybf{q},E_\Xcs) k_{\mu\nu}(\mybf{q},E_\Xcs) \, ,
\end{gather}
where $ k_{\mu\nu} $ is a known kinematic factor, $ G_F $ is the Fermi constant, and $ V_{cb} $ is the CKM matrix element related to the flavor-changing process $ b\rightarrow c $. The strong-interaction dynamics is encoded in the forward hadronic tensor $ W^{\mu\nu} $.

Through a Laplace transform, $ W^{\mu\nu} $ is related to a quantity 
that is accessible in lattice simulations:
\begin{gather}
    \scriptstyle  C_{J_\mu J_\nu}(\mybf{q} , t) \equiv \int \rmd^3 \mybf{x}\frac{ e^{i\mybf{q}\cdot\mybf{x}}}{2\MBs} \Braket{B_s| J_\mu^\dagger(\mybf{x},0) e^{-\hat{H}t} J_\nu(0) |B_s} = \int_{0}^{\infty} \rmd E_\Xcs e^{-tE_\Xcs} W^{\mu\nu}(\mybf{q} , E_{X_{cs}}) \, .\label{C_munu_def} 
\end{gather}
Inverting the Laplace transform to obtain the hadronic tensor from the lattice data is an ill-posed inverse problem. But for inclusive decays, what is most important is not $ W^{\mu\nu} $ itself, but its weighted integral over $ E_\Xcs $. Thus, expressing $ k_{\mu\nu} $ by polynomials of $ \exp(-E_{X_{cs}}) $, we can avoid this inverse problem to obtain estimates of the inclusive observables from $ C_{J_\mu J_\nu}(\mybf{q} , t) $. See \cite{Gambino:2020crt,Barone:2023iat,Kellermann:2023yec,Barone:2023tbl,Kellermann:2024zfy} for more details on the inclusive calculations.

On the other hand, the exclusive differential decay rate generally takes the following form
\begin{gather}
    \frac{\rmd \Gamma^{\mathrm{exc}}}{\rmd w} \propto G_F^2 \Abs{V_{cb}}^2 H\left(\mathcal{F}(w), \mathcal{K}(w,M_{\Xcs})\right) \, . 
\end{gather}
Here, $ \mathcal{F}(w) $'s are form factors to parameterize the relevant hadronic transition matrix elements; $ \mathcal{K}(w,M_{\Xcs}) $'s are known kinematical factors and $ H $ represents a collection of them. For exclusive processes, we introduce the recoil parameter $w\equiv v^\prime \cdot v$, with $v$ and $v^\prime$ the four-velocity of $B_s$ and $\Xcs$ respectively, and thus $ w = \sqrt{1+ \mybf{q}^2/M_{\Xcs}^2} $.

Usually, the non-perturbative form factors $ \mathcal{F}(w) $ are extracted from large time-separations of Euclidean three-point correlators, and only the ground-state (thus exclusive) form factors are obtained with meaningful precision. Here, we propose to extract them, along with the mass spectrum of the final states, from $ C_{J_\mu J_\nu}(\mybf{q} , t) $, which contains contributions from all possible final states. We notice that by inserting a complete set of final states, $ C_{J_\mu J_\nu}(\mybf{q} , t) $ becomes the sum of a series of exponentials
\begin{gather}
    C_{J_\mu J_\nu}(\mybf{q} , t) = \sum_{X_{cs}} \int\! \frac{d^3 \mybf{p}_\Xcs}{4E_\Xcs\MBs} \delta^{(3)}(\mybf{q} + \mybf{p}_{X_{cs}}) \Braket{B_s|J^\dagger_\mu(0)|X_{cs}} \Braket{X_{cs}| J_\nu(0) |B_s} e^{-E_{X_{cs}} t} \, . \label{eq:Cmunu_sum_Xcs}
\end{gather}
The prefactor of every exponential is nothing but the corresponding hadronic transition matrix element, which in turn is parameterized by form factors $ \mathcal{F}(w) $. Thus, we can in principle obtain exclusive information by performing a multiple-exponential fit of $ C_{J_\mu J_\nu}(\mybf{q} , t) $.

From \eq{\ref{C_munu_def}} and \eq{\ref{eq:Cmunu_sum_Xcs}}, it is clear that inclusive and exclusive differential decay rates are related to the same quantity $ C_{J_\mu J_\nu}(\mybf{q} , t) $. In this way, the saturation of the inclusive decay rate by exclusive channels should naturally be satisfied. This may provide a way to understand the tensions mentioned above, once precisely calculated on the lattice. Further, we also aspire to explore a new possibility to extract transition form factors into orbitally excited final states from lattice simulations with finite heavy-quark masses. Preliminary investigations, like that in \paper{\cite{Atoui:2013ksa}}, via three-point correlators are heavily obstructed by the complexity of constructing proper interpolating operators and the large noise.

\section{Decompositions of $ C_{J_\mu J_\nu} $ into different final states} \label{ss:decomposition_CJmuJnu}

Heavy-quark symmetry provides a baseline understanding of the spectrum and transition of heavy mesons. In the $ m_b, m_c \rightarrow \infty $ limit, the spectrum of the heavy-light mesons can be classified by the radial excitation and the total angular momentum of the light degrees of freedom. For each of the radial excitations, degenerate pairs can be found with the same total angular momentum and parity of the light degrees of freedom $ j^\mathcal{P} $. Here, we focus on the radial ground states of the $ D_s $ mesons. By coupling $ j^\mathcal{P}= (1/2)^- $ with the spin of the heavy quark $ s_c = 1/2 $, we obtain the $ S $-wave pair $ J^{\mathcal{P}} = (0^-,1^-) $, \ie{}, $ D_s $ and $ D_{s}^* $. The four lightest positive-parity states can also be classified into the $ P_{1/2} $ ($j^\mathcal{P}= (1/2)^+$) pair $ J^{\mathcal{P}} = (0^+,1^+) $, \ie{}, $ D_{s0}^* $ and $ D_{s1}^\prime $, and the $ P_{3/2} $ ($j^\mathcal{P}= (3/2)^+$) pair $ J^{\mathcal{P}} =  (1^+,2^+) $, \ie{}, $ D_{s1} $ and $ D_{s2} $. The mass splitting within each pair is proportional to $ 1/m_Q $ and thus vanishes in the heavy-quark limit.

At the zeroth order of HQET, the transitions from the $ B_s $ to the $ D_s $ mesons are described by three sets of form factors $ \xi^{(n)}(w) $, $ \tau_{1/2}^{(n)}(w) $ and $ \tau_{3/2}^{(n)}(w) $ for $ S $, $ P_{1/2} $ and $ P_{3/2} $ channels, respectively \cite{Leibovich:1997em,Isgur:1990jf}. Uraltsev proposed a sum rule among the $P$-wave form factors \cite{Uraltsev:2000ce}
\begin{gather}
    \scriptstyle  \sum_n \left(\Abs{\tau_{3/2}^{(n)}(1)}^2 - \Abs{\tau_{1/2}^{(n)}(1)}^2\right) = \frac{1}{4} \, .\label{Uraltsev_sum_rule}
\end{gather}
Here $ n $ is the quantum number of radial excitation. Naively, one expects the saturation of this sum rule from the radial ground state, \ie{}, $ \tau_{1/2}=\tau_{1/2}^{(0)} $ and $ \tau_{3/2}=\tau_{3/2}^{(0)} $, which suggests $ \tau_{3/2}(1) \gg \tau_{1/2}(1) $. This leads to the prediction that the branching ratios for the $(1/2)^+$ pairs should be suppressed compared to those of the $(3/2)^+$ pairs under the assumption that the form factors depend only mildly on $w$. This assumption should be verified by explicit numerical calculations.

\newcommand{\VanishZeroRecoil}[1]{{#1}}
Away from the heavy-quark limit, there are the following form factors to parameterize the semileptonic transitions of $ B_s $. For $S$-channel ($ j^\mathcal{P}=(1/2)^- $) final states, we have
\begin{align}
    \scriptstyle  \Braket{B_s,v| V_\mu |D_s,v^\prime} &= \left[ h_+ (v_\mu + v^\prime_\mu) + \VanishZeroRecoil{h_- (v_\mu - v^\prime_\mu)} \right] \sqrt{M_{B_s} M_{D_s}} \, , \label{eq:B_Vmu_D}\\
    \scriptstyle  \Braket{B_s,v| V_\mu |D^*_s,v^\prime,\epsilon} &= \left[\VanishZeroRecoil{h_V \varepsilon_{\mu\alpha\beta\gamma} \epsilon^{\alpha} v^{\prime,\beta} v^{\gamma}}\right] \sqrt{M_{B_s} M_{D^*_s}}  \, ,\\
    \scriptstyle  \Braket{B_s,v| A_\mu |D^*_s,v^\prime,\epsilon} &= i\left[(w + 1)h_{A1} \epsilon_\mu - \VanishZeroRecoil{(\epsilon\cdot v) \left(h_{A2} v_\mu + h_{A3} v^\prime_\mu\right)} \right] \sqrt{M_{B_s} M_{D^*_s}} \, .
\end{align}
For $ P_{1/2} $-channel ($ j^\mathcal{P}=(1/2)^+ $) final states, we have
\begin{align}
    \scriptstyle
    \Braket{B_s,v| A_\mu |D_{s0}^*,v^\prime} &= \left[ g_+ (v_\mu + v^\prime_\mu) + \VanishZeroRecoil{g_- (v_\mu - v^\prime_\mu)} \right] \sqrt{M_B M_{D_{s0}^*}} \, .
\end{align}
For $ P_{3/2} $-channel ($ j^\mathcal{P}=(3/2)^+ $) final states, we have
\begin{align}
    \scriptstyle  \Braket{B_s,v| V_\mu |D_{s1},v^\prime,\epsilon} &= \left[f_{V1} \epsilon_\mu + \VanishZeroRecoil{(\epsilon\cdot v) \left(f_{V2} v_\mu + f_{V3} v^\prime_\mu\right)} \right] \sqrt{M_{B_s} M_{D_{s1}}}  \, ,\\
    \scriptstyle  \Braket{B_s,v| A_\mu |D_{s1},v^\prime,\epsilon} &= -i \left[ \VanishZeroRecoil{ f_A \varepsilon_{\mu\alpha\beta\gamma} \epsilon^{\alpha} v^{\prime,\beta} v^{\gamma}} \right] \sqrt{M_{B_s} M_{D_{s1}}} \, . \label{eq:B_Amu_D1}
\end{align}
Here, $ \epsilon $ is the polarization vector for the spin-1 particle. $ \Braket{B_s,v| A_\mu |D_s,v^\prime} $ and $ \Braket{B_s,v| V_\mu |D_{s0}^*,v^\prime} $ vanish due to conservation of parity. All form factors $ h,g,f $ depend only on $w$. 

Due to limitations in the signal, we can only identify at most two exponentials from every $ C_{J_\mu J_\nu}(\mybf{q},t) $, which leads us to temporarily ignore the contributions from the highest spin state, \ie{} $ D_{s2} $. The vector final state in the $ P_{1/2} $-channel is also connected to four form factors $ g_{V1}, g_{V2}, g_{V3} $ and $ g_A $, but numerically they are around one order of magnitude smaller than the corresponding ones from the $ P_{3/2} $-channel (see discussions in \ssec{\ref{ss:numerics}}). 

By substituting \eqs{\ref{eq:B_Vmu_D}--\ref{eq:B_Amu_D1}} into \eq{\ref{eq:Cmunu_sum_Xcs}}, we obtain the decompositions of $ C_{J_\mu J_\nu}(\mybf{q},t) $ into a series of exponentials with their prefactors given by the corresponding form factors. We summarize the form-factor correspondence of $ C_{J_\mu J_\nu}(\mybf{q},t) $ in \tab{\ref{tab:dependence_Cmunu_FF}}. In this table, $ \perp $ and $ \parallel $ refer to the directions perpendicular and parallel to the three-momentum $ \mybf{q} $ and empty cells indicate that the $ C_{J_\mu J_\nu}(\mybf{q},t) $ in this row receives no contributions from the final state in this column.

\begin{table}
    \centering\renewcommand{\arraystretch}{1.0}
    \small
    \begin{tabular}{c || c : c | c | c }
        \hline
         & \multicolumn{2}{c|}{$S$-channel} & $P_{1/2}$-channel & $P_{3/2}$-channel \\\hdashline
         & $ D_s (0^-) $ & $ D_s^* (1^-) $ & $ D_{s0}^* (0^+) $ & $ D_{s1} (1^+) $ \\ \hline
        $ C_{V_0 V_0} $ & $ h_+, h_- $ & & & $ f_{V1}, f_{V2}, f_{V3} $ \\\hline
        $ C_{V_\parallel V_\parallel} $ & $ h_+, h_- $ & & &  $ f_{V1}, f_{V3} $ \\\hline
        $ C_{V_\perp V_\perp} $ & & $ h_V $ & &  $ f_{V1} $ \\\hline
        $ C_{V_0 V_\parallel} $ & $ h_+, h_- $ & & &  $ f_{V1}, f_{V2}, f_{V3} $ \\\hline
        $ C_{A_0 A_0} $ & & $ h_{A1}, h_{A2}, h_{A3} $ & $ g_+, g_- $ & \\\hline
        $ C_{A_\parallel A_\parallel} $ & & $ h_{A1}, h_{A3} $ & $ g_+, g_- $ & \\\hline
        $ C_{A_\perp A_\perp} $ & & $ h_{A1} $ & & $ f_A $ \\\hline
        $ C_{A_0 A_\parallel} $ & & $ h_{A1}, h_{A2}, h_{A3} $ & $ g_+, g_- $ & \\\hline
    \end{tabular}
    \caption{The dependence of $ C_{J_\mu J_\nu}(\mybf{q},t) $ on the form factors defined in \eqs{\ref{eq:B_Vmu_D}\textasciitilde\ref{eq:B_Amu_D1}}.\label{tab:dependence_Cmunu_FF}}
\end{table}

\section{Lattice setup}\label{ss:lattice}

In this study, we use a $ 24^3\times 64 $ lattice from RBC/UKQCD Collaboration with lattice spacing $ a^{-1}\approx 1.785\,\mathrm{GeV} $ \cite{Flynn:2023nhi}. 
We use DWF~\cite{Shamir:1993zy,Furman:1994ky}, M\"obius DWF~\cite{Brower:2012vk,Cho:2015ffa} and relativistic-heavy-quark action~\cite{Christ:2006us,Lin:2006ur} for the valence $s$, $c$, $b$ quarks respectively. Their masses are tuned such that the corresponding $ D_s $ and $ B_s $ have masses close to the physical ones. More details about the simulation can be found in \paper{\cite{Barone:2023tbl}} and references therein.

We start with the four-point correlators
\begin{gather}
    C^{S J_\mu J_\nu S} (t_{\snk},t_2,t_1,t_{\src},\mybf{q}) \propto \int \rmd^3 \mybf{x} \; e^{i\mybf{q}\cdot\mybf{x}} \Braket{0| \phi_{B_s}^S({t_{\snk}}) J^{\dagger}_\mu(\mybf{x},t_2) J_\nu(\mybf{0},t_1) \phi_{B_s}^{S\,\dagger}({t_{\src}}) |0} \, . \label{eq:def_four_pt}
\end{gather}
$ C_{J_\mu J_\nu}(\mybf{q},t) $ can then be obtained by taking the ratios between four-point correlators and two-point correlators
\begin{gather}
    C_{J_\mu J_\nu}(\mybf{q},t=t_2 - t_1) \propto \frac{C^{S J_\mu J_\nu S } (t_{\snk},t_2,t_1,t_{\src},\mybf{q})}{C^{S L} (t_\snk,t_2) C^{L S} (t_1,t_\src)} \, . \label{eq:from_four_pt_to_Cmunu}
\end{gather}
In this study, we fix $ t_\snk - t_\src = 20 $, $ t_2 - t_\src = 14 $ and vary $ t_1 $ from 0 to 14. $ L (S) $'s in \eqs{\ref{eq:def_four_pt}, \ref{eq:from_four_pt_to_Cmunu}} represent local (smearing) interpolating operators of $B_s$.

\section{Numerical results} \label{ss:numerics}

In the zero-recoil limit, only four correlators, $ C_{V_0 V_0} $, $ C_{A_\parallel A_\parallel} $, $ C_{A_0 A_0} $ and $ C_{V_\parallel V_\parallel} $ survive. Furthermore, due to parity conservation, contributions from different $ J^{\mathcal{P}} $ channels decouple from each other. Thus, at $w=1$, we have
\begin{align}
    \scriptstyle  C_{V_0 V_0}(t) &= e^{- E_{D_s} t } \left|h_+(1)\right|^2 \, , \\
    \scriptstyle  C_{\AparaApara}(t) &= e^{- E_{D_s^*} t } \left|h_{A1}(1)\right|^2 \, , \\
    \scriptstyle  C_{A_0 A_0}(t) &= e^{- E_{D_{s0}^*} t } \left|g_+(1)\right|^2 \, , \\
    \scriptstyle  C_{\VparaVpara}(t) &= e^{- E_{D_{s1}^\prime} t } \left|\frac{g_{V1}(1)}{2}\right|^2 + e^{- E_{D_{s1}} t } \left|\frac{f_{V1}(1)}{2}\right|^2 \approx e^{- E_{D_{s1}} t } \left|\frac{f_{V1}(1)}{2}\right|^2 \, , \label{eq:parameterization_C_ViVi_zero_recoil}
\end{align}
and all other $ C_{J_\mu J_\nu}(t) $ are zero. In \eq{\ref{eq:parameterization_C_ViVi_zero_recoil}}, we neglect the contribution from $ \left|g_{V1}(1)\right|^2 $ because by expanding them to the first order in the HQET \cite{Leibovich:1997em}, we have $ g_{V1}(1) \propto \left(\epsilon_c - 3 \epsilon_b\right) $ and $ f_{V1}(1) \propto \epsilon_c $. Here $ \epsilon_c \equiv 1/2m_c $ and $ \epsilon_b \equiv 1/2m_b $. The factor $ \left(\epsilon_c - 3 \epsilon_b\right) $ is numerically much smaller than $ \epsilon_c $. 

\begin{figure}
    \centering
    \includegraphics[width=0.4\textwidth]{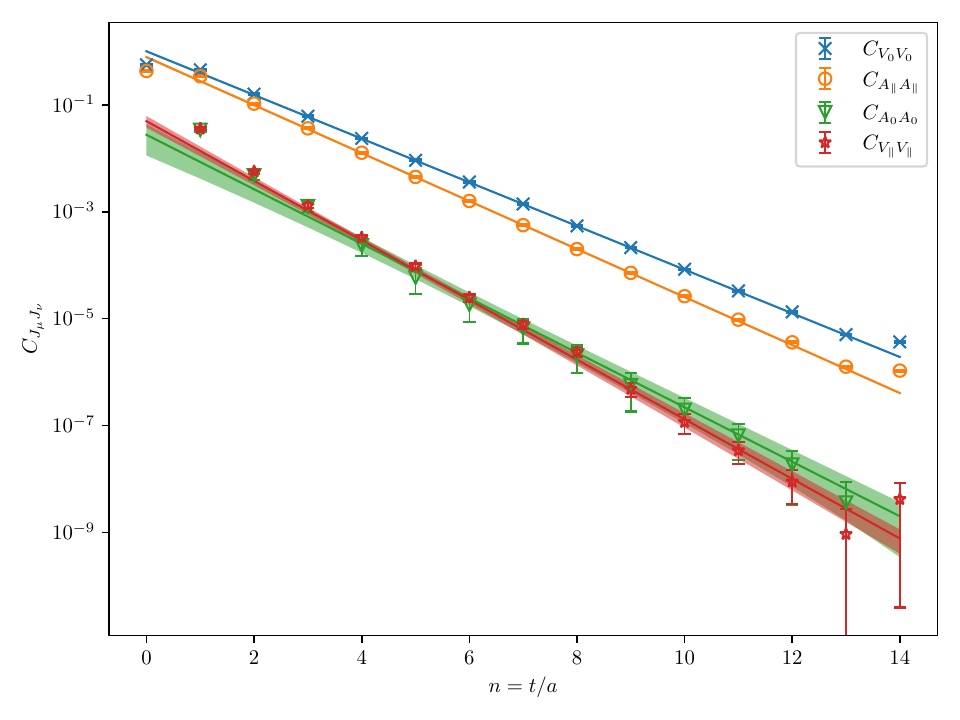}
    \includegraphics[width=0.4\textwidth]{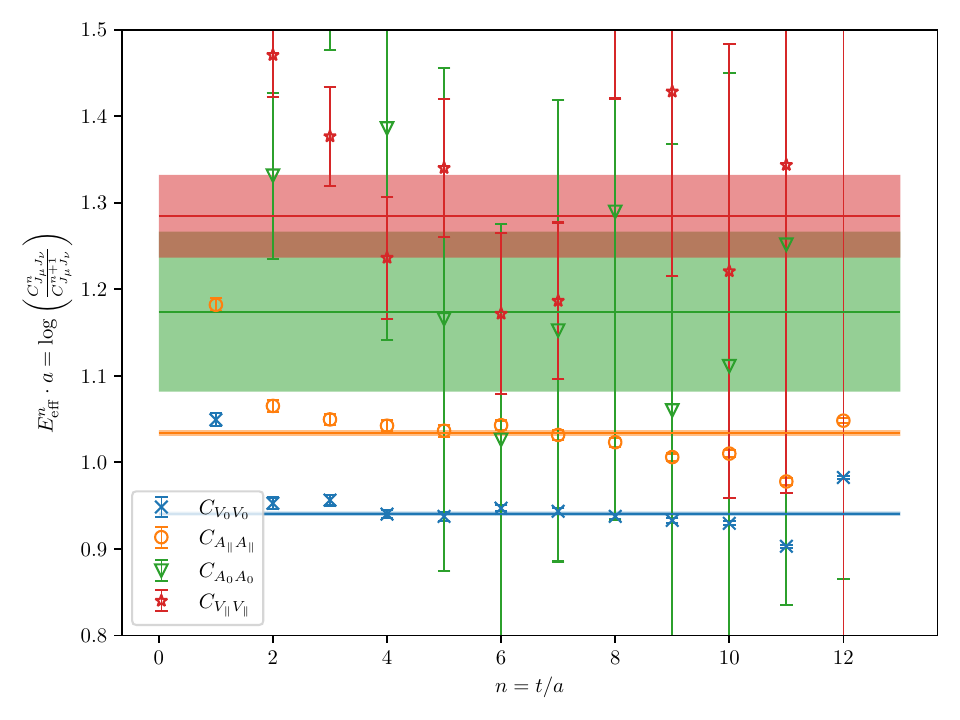}
    \caption{\label{fig:zero_recoil_limit_fitting_results} Single-exponential fit of the four non-vanishing correlators at the zero-recoil limit. In the left panel, we compare the fitted (straight lines with errors as bands) and the original (data points) correlators. In the right panel, the effective masses (disconnected points) and the extracted mass parameters (horizontal lines with errors as bands) are plotted together.}
\end{figure}

In \fig{\ref{fig:zero_recoil_limit_fitting_results}}, we show the results of the single-exponential fit for these four correlators (left panel). From the comparison of the correlators, we find that $ C_{V_0 V_0} $ and $ C_{\AparaApara} $, which correspond to the $ S $-wave states $ D_s $ and $ D_s^* $, show much stronger and cleaner signals than $ C_{\AzeroAzero} $ and $ C_{\VparaVpara} $. However, information about the $ P $-wave states $ D_{s0}^* $ and $ D_{s1} $ can still be extracted from the last two correlators. The mass hierarchy of those four states can be read from the comparisons of effective masses (right panel). Converting to physical units, we obtain $ M_{D_{s}^*} - M_{D_s} = 166.2\pm 4.9\,\MeV $, $ M_{D_{s0}^*} - M_{D_s} = 420\pm 160\,\MeV $ and $ M_{D_{s1}} - M_{D_s} = 613\pm 85\,\MeV $. Those values are consistent with PDG \cite{ParticleDataGroup:2024cfk} within errors.

Now, let us turn our attention to the form factors. From HQET expanded to the first order in $ 1/m_Q $, the form factors of the $ S $-wave final states at the zero-recoil limit are predicted to be
\begin{gather}
    h_+(1) \approx h_{A1}(1) \approx \xi(1) = 1 \, . \label{eq:hPlus_hA1_zero_recoil}
\end{gather}
We obtain $h_+(1) = 1.004\pm 0.021$ and $h_{A1}(1) = 0.885\pm0.022$, which implies significant (\textasciitilde$ 10\% $) contributions from the higher orders for $ D_{s}^* $ and less than a  percent contribution for $ D_s $. The higher order correction to $B_s\to D_s$ has the form $(\epsilon_c-\epsilon_b)^2$ while three separate combinations, $\epsilon_c^2$, $\epsilon_c\epsilon_b$, $\epsilon_b^2$, all appear for $B_s\to D_s^*$, signifying smaller higher order correction for $h_+(1)$. The results are also consistent with previous results summarized in FLAG 2024 \cite{FlavourLatticeAveragingGroupFLAG:2024oxs}.

For the $ P $-wave final states at the zero-recoil limit $w=1$, we obtain
$\Abs{g_{+}(1)} = 0.166\pm 0.049$ and $\Abs{f_{V1}(1)} = 0.445\pm 0.055$ from the lattice data. They are related to the Isgur-Wise form factors as
\begin{align}
    g_+(1) \approx -3 (\epsilon_c + \epsilon_b) \left( \overline{\Lambda^*_s} - \overline{\Lambda_s} \right) \tau_{1/2}(1) \, , \\
    f_{V1}(1) \approx -4\sqrt{2} \epsilon_c \left( \overline{\Lambda^\prime_s} - \overline{\Lambda_s} \right) \tau_{3/2}(1) \, .
\end{align}
To obtain an estimate for $\tau_{1/2}(1)$ and $\tau_{3/2}(1)$, we assume nominal values for the mass splittings among $ S $, $ P_{1/2} $ and $ P_{3/2} $ states in the $ m_Q\rightarrow\infty $ limit, $ \left( \overline{\Lambda^*_s} - \overline{\Lambda_s} \right) = 0.28\,\GeV $ and $ \left( \overline{\Lambda^\prime_s} - \overline{\Lambda_s} \right) = 0.41\,\GeV $, from a phenomenological analysis \cite{Bernlochner:2016bci}.
The quark masses to determine $\epsilon_b$ and $\epsilon_c$ are $m_b = 4.8\,\GeV$ and $m_c = 1.1\,\GeV$. We use a slightly smaller charm-quark mass compared to the phenomenological value \cite{Bernlochner:2016bci} in order to accommodate the lighter charm-quark mass in our lattice simulation. Our results are $\Abs{\tau_{1/2}(1)} = 0.35\pm 0.10$ and $\Abs{\tau_{3/2}(1)} = 0.423\pm 0.052$. 
 
At zero-recoil, form factor $ g_{V1} $ can also be expressed as
\begin{gather}
    g_{V1}(1) \approx 2 \left(\epsilon_c - 3 \epsilon_b\right) \left( \overline{\Lambda^*_s} - \overline{\Lambda_s} \right) \tau_{1/2}(1) \, .
\end{gather}
Using the value of $ \Abs{\tau_{1/2}(1)} $, we obtain $ \Abs{g_{V1}(1)}=0.0282\pm0.083 \ll \Abs{f_{V1}(1)} $, validating our choice to ignore the contribution of $ D_{s1}^\prime $ in \eq{\ref{eq:parameterization_C_ViVi_zero_recoil}}. We expect that this suppression is also valid at non-zero recoil, and thus also neglect the contributions from $ D_{s1}^\prime $ when $ \mybf{q}\neq\mybf{0} $ (see \tab{\ref{tab:dependence_Cmunu_FF}}). 

Moreover, we find $\left(\tau_{3/2}(1)\right)^2 - \left(\tau_{1/2}(1)\right)^2 = 0.053\pm0.079 $. The deficit compared to 1/4, see \eq{\ref{Uraltsev_sum_rule}}, may suggest significant contributions from the radial excitations. We notice, however, that the present study is performed using only one lattice ensemble with limited statistics, and the proper physical limits are still to be taken.

Finally we turn to the analysis at non-zero recoil. We perform multi-exponential fits for $C_{J_{\mu}J_{\nu}}(t)$'s at non-zero $\mybf{q}$.
After some simplifications, every four-point correlator is described by the summation of two exponentials with prefactors corresponding to the form factors (see \tab{\ref{tab:dependence_Cmunu_FF}})
\begin{gather}
        C_{J_{\mu}J_{\nu}}(t) = \mycal{A}_{J_{\mu}J_{\nu}}^{\mathrm{GS}} e^{-E_{J_{\mu}J_{\nu}}^{\mathrm{GS}} t } + \mycal{A}_{J_{\mu}J_{\nu}}^{\mathrm{EX1}} e^{-E_{J_{\mu}J_{\nu}}^{\mathrm{EX1}} t } \, . \label{eq:C_munu_GS_EX1}
\end{gather}
Here, $\mathrm{GS}$ stands for ground state and $\mathrm{EX1}$ stands for first excited state. Some of the $C_{J_{\mu}J_{\nu}}(t)$'s receive contributions from the same set of final states, and we perform simultaneous fits of them. We find that for both the ground state and the first excited state contributions, the equality $\mycal{A}_{V_0 V_\parallel} = \sqrt{\mycal{A}_{V_0 V_0}\times \mycal{A}_{V_\parallel V_\parallel}}$ holds. A similar relation can also be found for the prefactors of $C_{A_0 A_0}$, $C_{A_\parallel A_\parallel}$ and $C_{A_0 A_\parallel}$. We impose such equalities in our fits, and we introduce the lattice dispersion relation $E = \cosh^{-1} \left[ \cosh M + \sum_{i} \left( 1 - \cos p_i \right) \right]$ with the mass values obtained from the zero-recoil analysis to constrain the first exited state energy in \eq{\ref{eq:C_munu_GS_EX1}}.
From this fit, we could simultaneously obtain the contributions from both the $S$-channel final states and $P$-channel final states. But here we focus on the $P$-channel results.

We follow the approximation A used in \papers{\cite{Bernlochner:2016bci,Leibovich:1997em}} to express $\mycal{A}_{V_0 V_0}^{D_{s1}}$ and $\mycal{A}_{V_\parallel V_\parallel}^{D_{s1}}$ (we stress that $\mycal{A}_{V_0 V_\parallel}^{D_{s1}}$ is not a free parameter in our analysis due to the equality) using the zero-recoil values of the Isgur-Wise form factors $\tau(w) \equiv \sqrt{3} \tau_{3/2}(w) $ and its first derivative $\tau(w) = \tau_0 + \tau^\prime (w-1) + \dots$ as
\begin{align}
    \mycal{A}_{V_0 V_0}^{D_{s1}} &\approx \frac{1}{4w} \left\{ (w-1) \frac{16}{3} \left[ 1 + 2 \epsilon_c \left( \overline{\Lambda^\prime} - \overline{\Lambda} \right) \right] + 8 (w-1)^2 \right\} \left(\tau_0\right)^2 + (w-1)^2 \frac{8\tau_0}{3w} \tau^\prime \, , \label{eq:approxA_V0V0_Ds1}\\
    \mycal{A}_{V_\parallel V_\parallel}^{D_{s1}} &\approx \frac{1}{4w} \frac{8}{3} \left[ 2 \left( \overline{\Lambda^\prime} - \overline{\Lambda} \right) \epsilon_c + (w-1) \right]^2 \left(\tau_0\right)^2 \, . \label{eq:approxA_VparaVpara_Ds1}
\end{align}
Similarly, the contributions from $D_{s0}^*$ to $C_{A_0 A_0}$ and $C_{A_\parallel A_\parallel}$ can also be approximated using the zero-recoil values of the Isgur-Wise form factors $\zeta(w) \equiv 2 \tau_{1/2}(w) $ and its first derivative $\zeta(w) = \zeta_0 + \zeta^\prime (w-1) + \dots$ as
\begin{align}
    \mycal{A}_{A_0 A_0}^{D_{s0}^*} &\approx \frac{1}{4w} \left[ 3 \left( \overline{\Lambda^*} - \overline{\Lambda} \right) (\epsilon_c + \epsilon_b) +(w-1) \right]^2 \left(\zeta_0\right)^2 \, , \label{eq:approxA_A0A0_Ds0Star}\\
    \mycal{A}_{A_\parallel A_\parallel}^{D_{s0}^*} &\approx \frac{1}{4w} \left\{ 2 (w-1) \left[ 1 + 3 (\epsilon_c + \epsilon_b) \left( \overline{\Lambda^*} - \overline{\Lambda} \right) \right] + (w-1)^2 \right\} \left(\zeta_0\right)^2 + (w-1)^2 \frac{\zeta_0}{w} \zeta^\prime \, . \label{eq:approxA_AparaApara_Ds0Star}
\end{align}
The basic idea of approximation A is to treat $w-1$ as the same order as $ 1/m_Q $ and retain only the second order in $\mathcal{O}(w-1)\sim\mathcal{O}(1/m_Q)$. Thus, \eqs{\ref{eq:approxA_V0V0_Ds1}\textasciitilde\ref{eq:approxA_AparaApara_Ds0Star}} should only be valid at small $\mybf{q}^2$, or at $w$ close to $1$.

\begin{figure}
    \centering
    \includegraphics[width=0.45\textwidth]{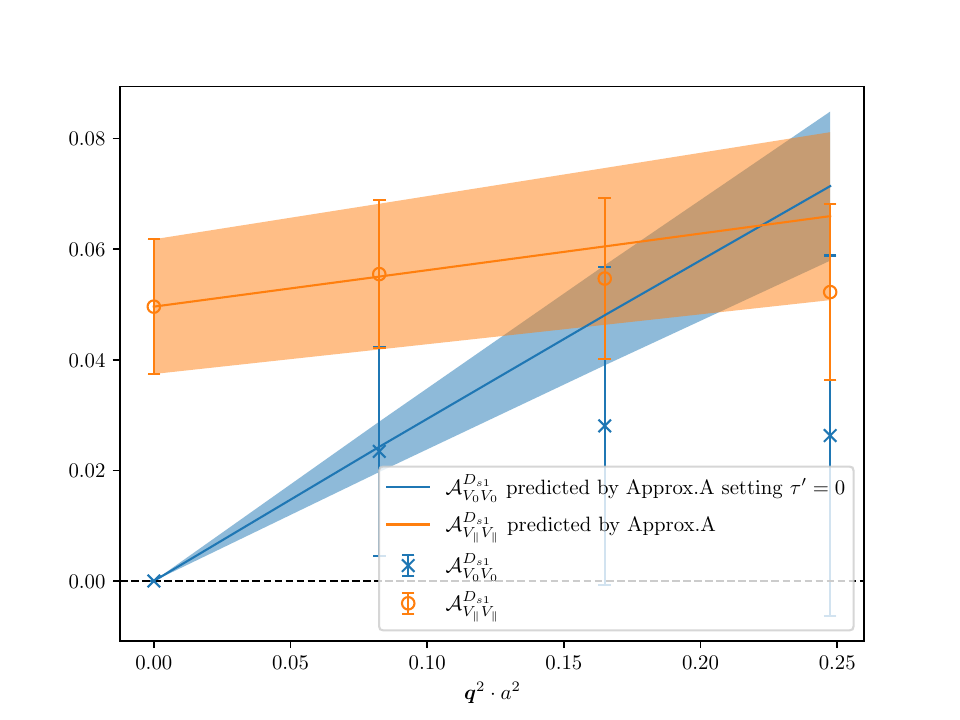}
    \includegraphics[width=0.45\textwidth]{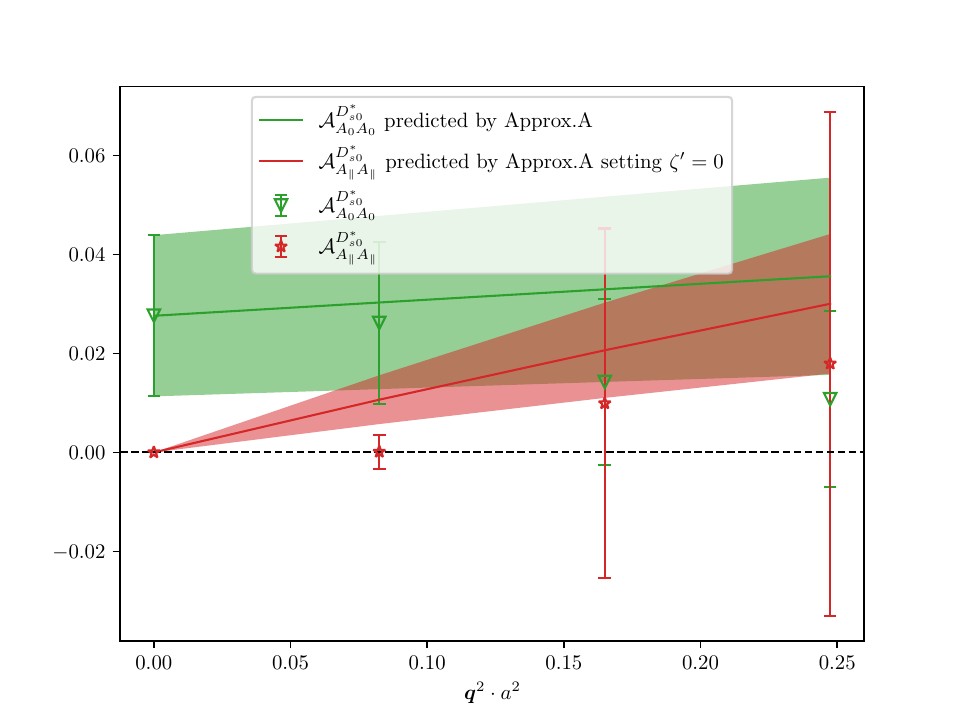}
    \caption{\label{fig:approxA} Four prefactors $\mycal{A}_{V_0 V_0}^{D_{s1}}$, $\mycal{A}_{V_\parallel V_\parallel}^{D_{s1}}$, $\mycal{A}_{A_0 A_0}^{D_{s0}^*}$, $\mycal{A}_{A_\parallel A_\parallel}^{D_{s9}^*}$ extracted from our fitting (data points) and those predicted by approximation A (\eqs{\ref{eq:approxA_V0V0_Ds1}\textasciitilde\ref{eq:approxA_AparaApara_Ds0Star}}, bands).}
\end{figure}

From the analysis of the correlators at zero recoil, we have $\tau_0 = \sqrt{3} \tau_{3/2}(1) = 0.73 \pm 0.09$ and $\zeta_0 = 2 \tau_{1/2}(1) = 0.71 \pm 0.21$, consistent with the phenomenological analysis from \paper{\cite{Bernlochner:2016bci}}. Thus, \eqs{\ref{eq:approxA_VparaVpara_Ds1}, \ref{eq:approxA_A0A0_Ds0Star}} can be used to check the validity of the approximation A and from \eqs{\ref{eq:approxA_V0V0_Ds1}, \ref{eq:approxA_AparaApara_Ds0Star}} we may extract the value of $\tau^\prime$ and $\zeta^\prime$. In \fig{\ref{fig:approxA}}, we plot the comparison between those four prefactors extracted from our fits to $C_{J_\mu J_\nu}$'s and the predictions from approximation A while setting $\tau^\prime$ and $\zeta^\prime$ to zero in \eqs{\ref{eq:approxA_V0V0_Ds1}, \ref{eq:approxA_AparaApara_Ds0Star}}. Consistency for $\mycal{A}_{V_\parallel V_\parallel}^{D_{s1}}$ and $\mycal{A}_{A_0 A_0}^{D_{s0}^*}$ can be readily observed at the smallest non-zero momentum while discrepancy increases with larger $\mybf{q}^2$. Thus, we consider it to be safer to extract $\tau^\prime$ and $\zeta^\prime$ only from the results at the smallest non-zero $\mybf{q}^2$. However, it can be observed that this is difficult since the extracted prefactors $\mycal{A}_{V_0 V_0}^{D_{s1}}$ and $\mycal{A}_{A_\parallel A_\parallel}^{D_{s0}^*}$ are dominated by error. Indeed, we obtain $\tau^\prime = - 1 \pm 17$ and $\zeta^\prime = - 18 \pm 10$. Such large errors prevent us from drawing any decisive conclusion. Closer investigations are demanded using finer lattice simulations and with better statistics.

\color{black}

\section{Discussions} \label{ss:discussions}

This work represents a pilot study of extracting information about the $ B_{(s)}\rightarrow D_{(s)} $ exclusive decays from the lattice four-point correlators $C_{J_{\mu}J_{\nu}}$. Our motivation is closely connected to the recent effort \cite{Kellermann:2024zfy,Barone:2023iat,Kellermann:2023yec,Barone:2023tbl} to extract inclusive decay width also from $C_{J_{\mu}J_{\nu}}$. By performing those two calculations based on the same set of lattice correlators, we open the possibility of simultaneous extractions of $V_{cb}$, and may alleviate the current tension about this CKM matrix element. However, this research requires careful multi-exponential fits of the lattice data and may demand larger source-sink separation and better statistics than that currently used to secure the extraction of excited-state contributions at non-zero recoil.

\section*{Acknowledgments}
This work used the DiRAC Extreme Scaling service at the University of Edinburgh, operated by the Edinburgh Parallel Computing Centre on behalf of the STFC DiRAC HPC Facility (www.dirac.ac.uk). This equipment was funded by BEIS capital funding via STFC capital grant ST/R00238X/1 and STFC DiRAC Operations grant ST/R001006/1. DiRAC is part of the National e-Infrastructure. The works of S.H. and T.K. are supported in part by JSPS KAKENHI Grant Numbers 22H00138, 22K21347 and 21H01085, and by the Post-K and Fugaku supercomputer project through the Joint Institute for Computational Fundamental Science (JICFuS).

\bibliographystyle{JHEP}
\bibliography{bib_file.bib}

\providecommand{\href}[2]{#2}\begingroup\raggedright\begin{thebibliography}{10}

\bibitem{HFLAV:2022esi}
{\scshape HFLAV} collaboration, \emph{{Averages of b-hadron, c-hadron, and \ensuremath{\tau}-lepton properties as of 2021}}, \href{https://doi.org/10.1103/PhysRevD.107.052008}{\emph{Phys. Rev. D} {\bfseries 107} (2023) 052008} [\href{https://arxiv.org/abs/2206.07501}{{\ttfamily 2206.07501}}].

\bibitem{Bigi:2007qp}
I.I.~Bigi, B.~Blossier, A.~Le~Yaouanc, L.~Oliver, O.~Pene, J.C.~Raynal et~al., \emph{{Memorino on the `1/2 versus 3/2 puzzle' in $\overline{B} \rightarrow l \overline{\nu} X_c$ - a year later and a bit wiser}}, \href{https://doi.org/10.1140/epjc/s10052-007-0425-1}{\emph{Eur. Phys. J. C} {\bfseries 52} (2007) 975} [\href{https://arxiv.org/abs/0708.1621}{{\ttfamily 0708.1621}}].

\bibitem{Hashimoto:2017wqo}
S.~Hashimoto, \emph{{Inclusive semi-leptonic B meson decay structure functions from lattice QCD}}, \href{https://doi.org/10.1093/ptep/ptx052}{\emph{PTEP} {\bfseries 2017} (2017) 053B03} [\href{https://arxiv.org/abs/1703.01881}{{\ttfamily 1703.01881}}].

\bibitem{Gambino:2020crt}
P.~Gambino and S.~Hashimoto, \emph{{Inclusive Semileptonic Decays from Lattice QCD}}, \href{https://doi.org/10.1103/PhysRevLett.125.032001}{\emph{Phys. Rev. Lett.} {\bfseries 125} (2020) 032001} [\href{https://arxiv.org/abs/2005.13730}{{\ttfamily 2005.13730}}].

\bibitem{Barone:2023iat}
A.~Barone, S.~Hashimoto, A.~J\"uttner, T.~Kaneko and R.~Kellermann, \emph{{Chebyshev and Backus-Gilbert reconstruction for inclusive semileptonic $B_{(s)}$-meson decays from Lattice QCD}}, \href{https://doi.org/10.22323/1.453.0236}{\emph{PoS} {\bfseries LATTICE2023} (2024) 236} [\href{https://arxiv.org/abs/2312.17401}{{\ttfamily 2312.17401}}].

\bibitem{Kellermann:2023yec}
R.~Kellermann, A.~Barone, S.~Hashimoto, A.~J\"uttner\ensuremath{\mathit{c}} and T.~Kaneko\ensuremath{\mathit{a}}, \emph{{Studies on finite-volume effects in the inclusive semileptonic decays of charmed mesons}}, \href{https://doi.org/10.22323/1.453.0272}{\emph{PoS} {\bfseries LATTICE2023} (2024) 272} [\href{https://arxiv.org/abs/2312.16442}{{\ttfamily 2312.16442}}].

\bibitem{Barone:2023tbl}
A.~Barone, S.~Hashimoto, A.~J\"uttner, T.~Kaneko and R.~Kellermann, \emph{{Approaches to inclusive semileptonic B$_{(s)}$-meson decays from Lattice QCD}}, \href{https://doi.org/10.1007/JHEP07(2023)145}{\emph{JHEP} {\bfseries 07} (2023) 145} [\href{https://arxiv.org/abs/2305.14092}{{\ttfamily 2305.14092}}].

\bibitem{Kellermann:2024zfy}
R.~Kellermann, A.~Barone, S.~Hashimoto, A.~J\"uttner and T.~Kaneko, \emph{{Updates on inclusive charmed and bottomed meson decays from the lattice}},  in \emph{{12th International Workshop on the CKM Unitarity Triangle}} \href{https://arxiv.org/abs/2405.06152}{{\ttfamily 2405.06152}}.

\bibitem{Atoui:2013ksa}
M.~Atoui, B.~Blossier, V.~Mor\'enas, O.~P\`ene and K.~Petrov, \emph{{Semileptonic $B \to D^{**}$ decays in Lattice QCD : a feasibility study and first results}}, \href{https://doi.org/10.1140/epjc/s10052-015-3585-4}{\emph{Eur. Phys. J. C} {\bfseries 75} (2015) 376} [\href{https://arxiv.org/abs/1312.2914}{{\ttfamily 1312.2914}}].

\bibitem{Leibovich:1997em}
A.K.~Leibovich, Z.~Ligeti, I.W.~Stewart and M.B.~Wise, \emph{{Semileptonic B decays to excited charmed mesons}}, \href{https://doi.org/10.1103/PhysRevD.57.308}{\emph{Phys. Rev. D} {\bfseries 57} (1998) 308} [\href{https://arxiv.org/abs/hep-ph/9705467}{{\ttfamily hep-ph/9705467}}].

\bibitem{Isgur:1990jf}
N.~Isgur and M.B.~Wise, \emph{{Excited charm mesons in semileptonic anti-B decay and their contributions to a Bjorken sum rule}}, \href{https://doi.org/10.1103/PhysRevD.43.819}{\emph{Phys. Rev. D} {\bfseries 43} (1991) 819}.

\bibitem{Uraltsev:2000ce}
N.~Uraltsev, \emph{{New exact heavy quark sum rules}}, \href{https://doi.org/10.1016/S0370-2693(01)00110-1}{\emph{Phys. Lett. B} {\bfseries 501} (2001) 86} [\href{https://arxiv.org/abs/hep-ph/0011124}{{\ttfamily hep-ph/0011124}}].

\bibitem{Flynn:2023nhi}
{\scshape RBC/UKQCD} collaboration, \emph{{Exclusive semileptonic Bs\textrightarrow{}K\ensuremath{\ell}\ensuremath{\nu} decays on the lattice}}, \href{https://doi.org/10.1103/PhysRevD.107.114512}{\emph{Phys. Rev. D} {\bfseries 107} (2023) 114512} [\href{https://arxiv.org/abs/2303.11280}{{\ttfamily 2303.11280}}].

\bibitem{Shamir:1993zy}
Y.~Shamir, \emph{{Chiral fermions from lattice boundaries}}, \href{https://doi.org/10.1016/0550-3213(93)90162-I}{\emph{Nucl. Phys. B} {\bfseries 406} (1993) 90} [\href{https://arxiv.org/abs/hep-lat/9303005}{{\ttfamily hep-lat/9303005}}].

\bibitem{Furman:1994ky}
V.~Furman and Y.~Shamir, \emph{{Axial symmetries in lattice QCD with Kaplan fermions}}, \href{https://doi.org/10.1016/0550-3213(95)00031-M}{\emph{Nucl. Phys. B} {\bfseries 439} (1995) 54} [\href{https://arxiv.org/abs/hep-lat/9405004}{{\ttfamily hep-lat/9405004}}].

\bibitem{Brower:2012vk}
R.C.~Brower, H.~Neff and K.~Orginos, \emph{{The M\"obius domain wall fermion algorithm}}, \href{https://doi.org/10.1016/j.cpc.2017.01.024}{\emph{Comput. Phys. Commun.} {\bfseries 220} (2017) 1} [\href{https://arxiv.org/abs/1206.5214}{{\ttfamily 1206.5214}}].

\bibitem{Cho:2015ffa}
Y.-G.~Cho, S.~Hashimoto, A.~J\"uttner, T.~Kaneko, M.~Marinkovic, J.-I.~Noaki et~al., \emph{{Improved lattice fermion action for heavy quarks}}, \href{https://doi.org/10.1007/JHEP05(2015)072}{\emph{JHEP} {\bfseries 05} (2015) 072} [\href{https://arxiv.org/abs/1504.01630}{{\ttfamily 1504.01630}}].

\bibitem{Christ:2006us}
N.H.~Christ, M.~Li and H.-W.~Lin, \emph{{Relativistic Heavy Quark Effective Action}}, \href{https://doi.org/10.1103/PhysRevD.76.074505}{\emph{Phys. Rev. D} {\bfseries 76} (2007) 074505} [\href{https://arxiv.org/abs/hep-lat/0608006}{{\ttfamily hep-lat/0608006}}].

\bibitem{Lin:2006ur}
H.-W.~Lin and N.~Christ, \emph{{Non-perturbatively Determined Relativistic Heavy Quark Action}}, \href{https://doi.org/10.1103/PhysRevD.76.074506}{\emph{Phys. Rev. D} {\bfseries 76} (2007) 074506} [\href{https://arxiv.org/abs/hep-lat/0608005}{{\ttfamily hep-lat/0608005}}].

\bibitem{ParticleDataGroup:2024cfk}
{\scshape Particle Data Group} collaboration, \emph{{Review of particle physics}}, \href{https://doi.org/10.1103/PhysRevD.110.030001}{\emph{Phys. Rev. D} {\bfseries 110} (2024) 030001}.

\bibitem{FlavourLatticeAveragingGroupFLAG:2024oxs}
{\scshape Flavour Lattice Averaging Group (FLAG)} collaboration, \emph{{FLAG Review 2024}},  \href{https://arxiv.org/abs/2411.04268}{{\ttfamily 2411.04268}}.

\bibitem{Bernlochner:2016bci}
F.U.~Bernlochner and Z.~Ligeti, \emph{{Semileptonic $B_{(s)}$ decays to excited charmed mesons with $e,\mu,\tau$ and searching for new physics with $R(D^{**})$}}, \href{https://doi.org/10.1103/PhysRevD.95.014022}{\emph{Phys. Rev. D} {\bfseries 95} (2017) 014022} [\href{https://arxiv.org/abs/1606.09300}{{\ttfamily 1606.09300}}].

\end{thebibliography}\endgroup

\end{document}